\RequirePackage{fix-cm}
\documentclass[twocolumn,epjc3]{svjour3}  

\smartqed  
\usepackage{color}
\usepackage{xcolor}
\definecolor{bred}{rgb}{0.8, 0.0, 0.0}
\definecolor{pblue}{rgb}{0.2, 0.2, 0.6}
\definecolor{ao}{rgb}{0.0, 0.5, 0.0}
\definecolor{carmine}{rgb}{0.59, 0.0, 0.09}

\newcommand{\CEvNS}{CE$\nu$NS}
\newcommand{\conusplus}{CONUS\texttt{+}}

\newcommand\Tstrut{\rule{0pt}{2.5ex}}         
\newcommand\Bstrut{\rule[-1.4ex]{0pt}{0pt}}   
\newcommand{\inches}{\ensuremath{{}^{\prime\prime}}}

\usepackage{upgreek}
\usepackage{subcaption}
\usepackage{url}
\usepackage{gensymb}
\usepackage{tabularx}
\usepackage{graphicx}
\usepackage{amsmath}
\usepackage{amssymb}
\usepackage{hyperref}
\usepackage{cite}
\usepackage{arydshln}

\usepackage{hyperref}
\hypersetup{
    colorlinks = true,
    linkbordercolor = {white},
    linkcolor = bred,
    citecolor = pblue,
    urlcolor = pblue
}

\usepackage[switch]{lineno} 

\begin{document}\sloppy

\title{The \conusplus~experiment }
\titlerunning{The \conusplus~experiment}

\author{The CONUS Collaboration: N.~Ackermann\thanksref{MPIK}, S.~Armbruster\thanksref{MPIK}, H.~Bonet\thanksref{MPIK}, C.~Buck\thanksref{MPIK}, K.~F\"{u}lber\thanksref{KBR}, J.~Hakenm\"{u}ller\thanksref{MPIK, Duke}, J.~Hempfling\thanksref{MPIK},  G.~Heusser\thanksref{MPIK}, M.~Lindner\thanksref{MPIK}, W.~Maneschg\thanksref{MPIK}, K.~Ni\thanksref{MPIK}, M.~Rank\thanksref{KKL}, T.~Rink\thanksref{MPIK}, E.~S\'{a}nchez Garc\'{i}a\thanksref{MPIK}, I.~Stalder\thanksref{KKL}, H.~Strecker\thanksref{MPIK}, R.~Wink\thanksref{KBR}, J.~Woenckhaus\thanksref{KKL,PSI} 
}

\authorrunning{N. Ackermann et al.}
\institute{Max-Planck-Institut f\"ur Kernphysik, Saupfercheckweg 1, 69117 Heidelberg, Germany \label{MPIK}  \and PreussenElektra GmbH, Kernkraftwerk Brokdorf, Osterende, 25576 Brokdorf, Germany  \label{KBR} \and Kernkraftwerk Leibstadt AG, 5325 Leibstadt, Switzerland  \label{KKL} \and \emph{Present Address:} Duke University, NC 27708, USA\label{Duke} \and \emph{Present Address:} Paul Scherrer Institut, Forschungsstrasse 111, 5232 Villigen, Switzerland\label{PSI} \vspace*{0.2cm}\vspace*{0.2cm} 
\\ E-mail address: \href{mailto:conus.eb@mpi-hd.mpg.de}{conus.eb@mpi-hd.mpg.de}
\\ E-mail address: \href{mailto:esanchez@mpi-hd.mpg.de}{esanchez@mpi-hd.mpg.de}}

\date{\today}

\maketitle

\begin{abstract}
  The \conusplus~experiment aims to detect coherent elastic neutrino-nucleus scattering (\CEvNS) of reactor antineutrinos on germanium nuclei in the fully coherent regime, continuing on this way the CONUS physics program started at the Brokdorf nuclear power plant, Germany. The \conusplus~setup is installed in the nuclear power plant in Leibstadt, Switzerland, at a distance of 20.7 m from the 3.6 GW thermal power reactor core. The \CEvNS~signature will be measured with the same four point-contact high-purity germanium (HPGe) detectors produced for the former experiment, however refurbished and with optimized low energy thresholds of about 160~eV$_{ee}$. To suppress the background in the \conusplus~detectors, the passive and active layers of the original CONUS shield were modified such to fit better to the significantly changed background conditions at the new experimental location. New data acquisition and monitoring systems were developed. A direct network connection between the experiment and the  Max-Planck-Institut f\"{u}r Kernphysik (MPIK) makes it  possible to control and monitor data acquisition in real time. The impact of all these modifications is discussed with particular emphasis on the resulting \CEvNS~signal prediction for the first data collection phase of \conusplus. Prospects of the planned upgrade in a second phase integrating new larger HPGe detectors are also discussed. 

\end{abstract}

\section{Introduction}
\label{intro}
 

Coherent elastic neutrino-nucleus scattering (\CEvNS) is a fundamental interaction within the framework of the Standard Model (SM) of particle physics, which was predicted in 1974~\cite{Freedman:1973yd,Kopeliovich:1974mv}. One distinctive aspect of this process lies in its notably higher cross section when compared to other neutrino interaction channels, such as inverse beta decay (IBD) or elastic neutrino-electron scattering (E$\nu$S)\cite{Barbeu:2023}. This opens up the possibility to build compact neutrino detectors that can, for instance, be installed close to nuclear power plants for reactor safeguarding and monitoring~\cite{vonRaesfeld:2021gxl,CHANDLER:2022gvg}. Additionally, as the interaction occurs at small momentum transfer, the weak form factor is approaching unity. Further, \CEvNS~research provides means to explore a diverse array of elementary particle physics processes, both within the Standard Model (SM) and beyond it (BSM)~(e.g. \cite{CONUS:2021dwh,CONUS:2022qbb}). 

 The \CEvNS~process was measured for the first time by the COHERENT Collaboration at a spallation neutron source, where neutrinos are generated through pion decays at rest ($\pi$-DAR)~\cite{Coherent:2017,COHERENT:2020iec,Adamski:2024yqt}. An alternative source of neutrinos for \CEvNS~investigations are nuclear reactors. In contrast to $\pi$-DAR sources, reactor antineutrinos exhibit lower energies below 12 MeV, guaranteeing a \CEvNS~detection in the fully coherent regime. Currently, there is a large effort worldwide to detect \CEvNS~at reactors~\cite{Aguilar-Arevalo:2024dln,Ackermann:2024kxo,MINER:2016igy,Colaresi:2022obx,NEON:2022hbk,NUCLEUS:2022zti,nGeN:2022uje,Akimov:2022xvr,Yang:2024exl,Cai:2024bpv,Ricochet:2023nvt,Karmakar:2024ydi}.

The CONUS experiment was operated from April 2018 to December 2022 at the nuclear power plant in Brokdorf (Kernkraftwerk Brokdorf; KBR), Germany, providing the best limit for \CEvNS~detection from a reactor to date, within a factor 1.6 over the prediction of the Standard Model~\cite{Ackermann:2024kxo}. Four p-type point-contact (PPC) high purity germanium (HPGe) detectors named C1 to C4, which were cooled with electrical cryocoolers around 85~K, were used as targets. They had a total active mass of 3.74~kg and an energy threshold of 210~eV$_{ee}$. The unit eV$_{ee}$ (electron equivalent energy) refers to the ionization energy observed in such detectors. Multiple layers of passive shield and an active muon veto around these detectors reduced the background level by four orders of magnitude to $\sim$10~counts~kg$^{-1}$d$^{-1}$~\cite{Bonet:2020ntx} in the [0.4, 1]~keV$_{ee}$ energy region. However, the KBR reactor finished operation by the end of 2021, which prompted the search for a new location for the continuation of the experiment. 

A new location for the successor experiment \conusplus~was found at the nuclear power plant in Leibstadt (Kernkraftwerk Leibstadt; KKL), Switzerland, and the setup was installed during summer 2023. The first phase of \conusplus~foresees the usage of four out of five refurbished HPGe detectors of CONUS (C2-C4, and the auxiliary C5), that have improved energy thresholds and trigger efficiencies in the low energy region compared to the original version. To match the new background conditions at KKL \cite{Sanchez:2024} the shield configuration was modified, too. In particular, the weaker reactor correlated $\gamma$-radiation and the smaller overburden at the new location compared to KBR, motivated the replacement of a lead layer by a second active muon veto system. In addition, a network connection to the outside of the containment area was established, allowing for a continuous monitoring of the experiment. This helps to maximize stability of data collection and exposure time.

This article reports about the design and the expected performance of the \conusplus~ experiment.  The description of the experimental location is detailed in Sec.\,\ref{sec:2}. The upgrade of the experimental setup from adaptations in the shield design to optimisation of the germanium detectors is described in Sec.\,\ref{sec:3}. The commissioning of the \conusplus~experiment is reported in Sec.\,\ref{sec:4}. To conclude, the \CEvNS~signal prediction for the current first phase of \conusplus~and for the planned upgrade in a second phase are discussed in Sec.\,\ref{sec:5}.

\section{\conusplus~location at KKL reactor}
\label{sec:2}


Since November 2023, the \conusplus~detector setup has been operational at the KKL nuclear power plant. This boiling water reactor consists of a single unit power station with a reactor core of 648~fuel assemblies. The maximum thermal power amounts to 3.6~GW~\cite{KKL2024}. The absolute thermal power is determined by a heat balance around the reactor. This information is available in intervals of 1 hour or less, with an uncertainty below 2\%. The reactor is operated at a high duty cycle and most of the time at its maximum thermal power. During an annual outage of about one month, called in the following reactor off period, general maintenance is done and part of the fuel elements are exchanged.

\begin{figure}
    \centering
    \includegraphics[width=0.47\textwidth]{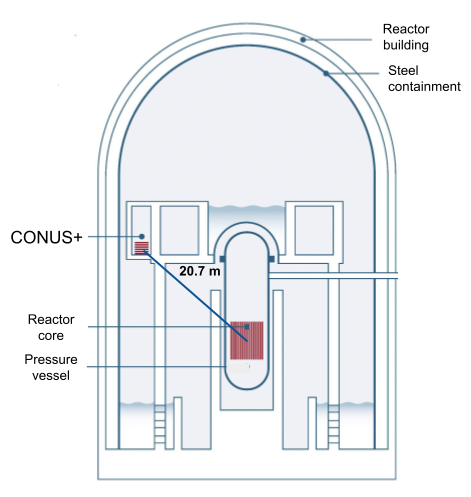}
    \caption{Scheme of the \conusplus~location inside the KKL containment. The roof of the ZA28R027 room and the reactor building provide an average overburden of 7.4~m~w.e.}
    \label{power_plant}
\end{figure}

The \conusplus~experiment is located in the room ZA28R027 inside the containment of KKL. The location of this room is shown in figure~\ref{power_plant}. It is situated at 25.2~m over the ground surface and at a horizontal distance of 12.45~m from the reactor central axis. The reactor core is situated at 9.9~m above ground and it has an extension of 6~m in diameter and 3.8~m in height. The four ultra-low energy threshold, high-purity germanium (HPGe) detectors of \conusplus~are positioned at (20.7~$\pm$~0.1)~m from the reactor core center. This leads to an approximately pointlike antineutrino source with a flux of $1.45\cdot 10^{13}$~s$^{-1}$~cm$^{-2}$ at the \conusplus~site~\cite{Kopeikin:2003gu}.

The concrete structure of KKL provides an average overburden of 7.4~m of water equivalent (m w.e.), which is sufficient to shield against the hadronic component of cosmic rays at surface, reducing the neutron flux by almost two orders of magnitude~\cite{HAUSSER1993223}. Two concrete walls that are part of the biological shield surrounding the pressure vessel reduce the reactor radiation by many orders of magnitude reaching an average radiation dose below 1~$\mu$Sv h$^{-1}$ in the \conusplus~area.

\section{The \conusplus~upgrades}
\label{sec:3}

\subsection{Passive and active shield upgrade}
\label{sec:3.1}

A good understanding of all background contributions is crucial for the success of a rare-event-physics experiment like \conusplus. Cosmic muons can produce electromagnetic cascades and neutrons within the materials that they traverse, particularly if the materials have a high proton number (Z) as lead (Pb) or steel-reinforced structures. Additionally, events correlated with the reactor thermal power are troublesome, as they can mimic the predicted  \CEvNS~interactions. For this reason, an extensive background characterization campaign was conducted during reactor on and off periods at the \conusplus~experimental location~\cite{Sanchez:2024}. 


On-site measurements revealed a highly thermalized and thermal power correlated neutron field larger than at KBR~\cite{Hakenmuller:2019ecb}, but still remaining a sub-dominant background component. The high energy ($\geq$~3~MeV) $\gamma$-ray background correlated with the thermal power was found to be reduced by factor of $\sim$26 compared to KBR. This is mainly due to the larger distance to the reactor water cooling cycle in which $\gamma$-emitting $^{16}$N decays take place. The cosmic muon rate of (107~$\pm$~3)~Hz~cm$^{-2}$ measured at KKL is 2.3~times larger than at KBR due to the smaller average overburden of (7.4~$\pm$~0.1)~m~w.e. More details about the new background conditions at KKL are discussed in~\cite{Sanchez:2024}.



To cope with the changed background conditions at the new experimental site and to achieve similar background levels as in the former CONUS experiment, the onion-like CONUS shield setup~\cite{Hakenmuller:2019ecb, Bonet:2021wjw} was modified for \conusplus. In particular, the middle out of the five lead (Pb) layers was replaced with an additional muon veto. This had a fourfold motivation. First, the removal of 5 cm of Pb is acceptable, since the 10-fold decrease in high-energy $\gamma$-ray attenuation is by far compensated by the 26-fold reduced $\gamma$-radiation at KKL compared to KBR. Second, the prompt signals induced by the enhanced muon flux at KKL require an improved muon veto rejection efficiency compared to KBR due to the smaller overburden. The production of delayed muon-induced neutrons in Pb is by 2.3 times enhanced at KKL compared to KBR~\cite{Sanchez:2024}, which has already represented one of the main backgrounds in CONUS~\cite{Bonet:2021wjw}; the enhanced muon veto efficiency is beneficial also against such neutrons. Third, the additional content of hydrogen in the plastic scintillators is also useful to further protect the HPGe detectors against external neutrons. And fourth, the exchange of the Pb layer with the second muon veto leads to a shield mass reduction of 1.6~tons, which has a positive impact on the load transfer of the static calculations.

\begin{figure}
    \centering
    \includegraphics[width=0.5\textwidth]{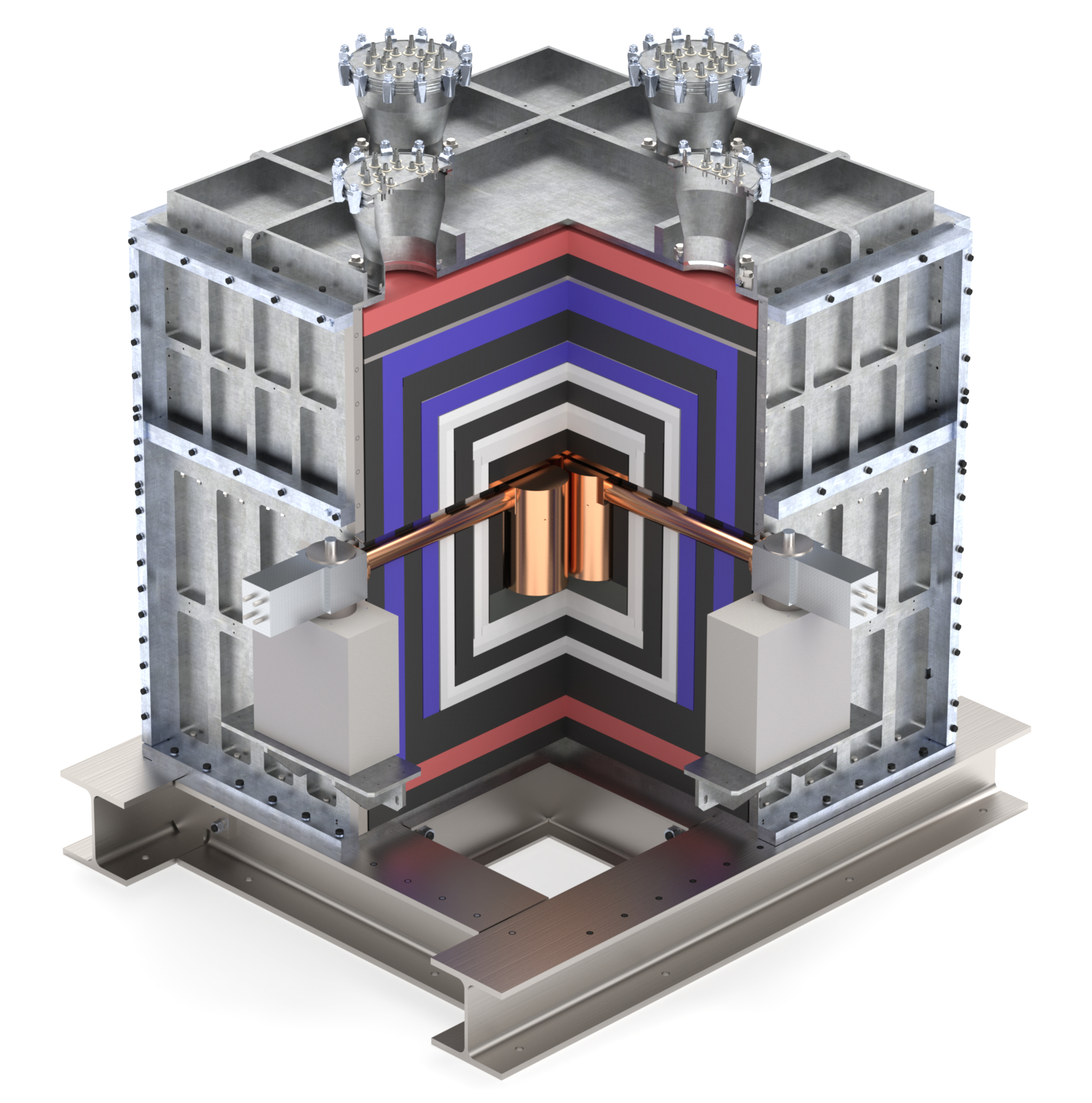}
    \caption{\conusplus~setup. The shield consists of layers made of stainless steel (silver), PE (red), Pb (black) and B-doped PE (white). Further it embeds two layers of plastic scintillator plates equipped with PMTs (blue) and used as a double muon veto system. In the center, four HPGe detectors are embedded in ultra-low background copper (Cu) cryostats and connected to electrically powered water-based cryocooler units.}
    \label{conus_shield}
\end{figure}

The finally implemented \conusplus~shield is depicted in figure~\ref{conus_shield}. It has four layers of Pb (black) providing a total thickness of 20 cm in all directions (except for the unchanged bottom with 25 cm of Pb) and with increasing radiopurity towards the detector chamber. The innermost layer consists of radiopure Pb with an average $^{210}$Pb concentration of less than 1.1~Bq kg$^{-1}$. Two layers of scintillation plates (blue) of type EJ-200~\cite{EJ200_ref} are installed as an active muon veto, accounting for a total of 18 individual scintillator plates. Each of them is equipped with two photomultiplier tubes (PMTs), aside for the top-side modules that have four PMTs. For the outer layer PMTs of type Hamamatsu R11265 U-2000~\cite{R11265U-200_ref} were used, while for the inner layer Hamamatsu R8520-4060~\cite{R8520-406_ref} were installed. Two polyethylene plates (PE) (red) and two layers of boron-doped (B-doped) PE (white) are installed. Together with the hydrogen-rich organic scintillator plates, they provide an efficient shield against any source of neutrons. The setup is enclosed by a steel frame (silver), which has been reinforced compared to CONUS at KBR in order to fulfill the stricter earthquake requirements of the new site at KKL. Additionally, radon (Rn) diffusion is reduced by using radon-seal tape at the edges of the frame. The whole shield has a total volume of 1.65 m$^{3}$ and a total mass of 9.94~tons, wherein the mass of the substructure is included.

The impact of removing 5~out of 25~cm of Pb was studied in detail with Monte-Carlo simulations using the framework MaGe~\cite{Boswell:2010mr}. This code is well validated for electromagnetic and hadronic processes at the keV-to-MeV scale. The $\gamma$-radiation was simulated according to the performed measurements during the background characterization of the experimental location, paying special attention to the high-energy $\gamma$-lines with the strongest contributions ($^{53}$Fe, $^{56}$Fe and $^{63}$Cu). Due to computational time limitation an upper limit of $<$~0.5~counts~kg$^{-1}$~d$^{-1}$ in the [0.4, 1]~keV$_{ee}$ energy region was obtained, which is subdominant compared to the previous background levels in CONUS~\cite{Bonet:2021wjw}. Radiopure materials were carefully selected for the new inner muon veto. The scintillator plates were screened at Max-Planck-Institut f\"{u}r Kernphysik (MPIK), and the PMTs and their electronic bases were previously used and tested by the XENON100 collaboration~\cite{Aprile:2011ru}. The effect of the new material over the total background was simulated, achieving an upper limit of $<$~2.1~counts~kg$^{-1}$~yr$^{-1}$ in the [0.4, 1]~keV$_{ee}$ energy region. In this way, a negligible negative impact of the shield modifications over the background budget was probed. The new double muon veto system at KKL provides a background rejection efficiency over 98\%, improving the previous KBR value of 97\%~\cite{Bonet:2020ntx}, thus compensating for the increased muon flux due to the smaller overburden. Overall, the \conusplus~shield keeps the previous CONUS background suppression of four orders of magnitude~\cite{Bonet:2021wjw}. 

The new inner muon veto was tested in the MPIK underground low-level laboratory (LLL) before its installation at KKL. The individual PMTs were tested before their assembly using a dark box, selecting the ones with better performance and similar dark current rates. In order to ensure that both muon veto systems have similar performance, the same company (Scionix) provided the plastic scintillator and integrated the PMTs. After their installation, the PMTs were tested again at MPIK and calibrated using dark current events present in the tail of scintillation pulses after several $\mu$s. The gain variation in the [750, 900]~V range was studied, finding a linear behavior and setting a similar gain for each PMT. The light yield of each scintillator plate was estimated with a $^{60}$Co source. A light yield of ($174\pm9$)~p.e. keV$^{-1}$ -- with p.e. being photoelectrons -- was measured for the top plate and an average of ($177\pm7$)~p.e. keV$^{-1}$ and ($70\pm3$)~p.e. keV$^{-1}$ for the lateral bottom and top plates, respectively. Differences of less than 5\% between plates of the same dimension were found. The light yield variations between the scintillator configurations are due to the different plate dimensions and number of PMTs installed. Light propagation simulations were performed using the parameters reported in~\cite{EJ200_ref} and a reflectivity of 99\% at 400~nm for inner plastic scintillator reflector. Good agreement was achieved with differences of less than 20\%.       

\begin{figure}
    \centering
    \includegraphics[width=0.48\textwidth]{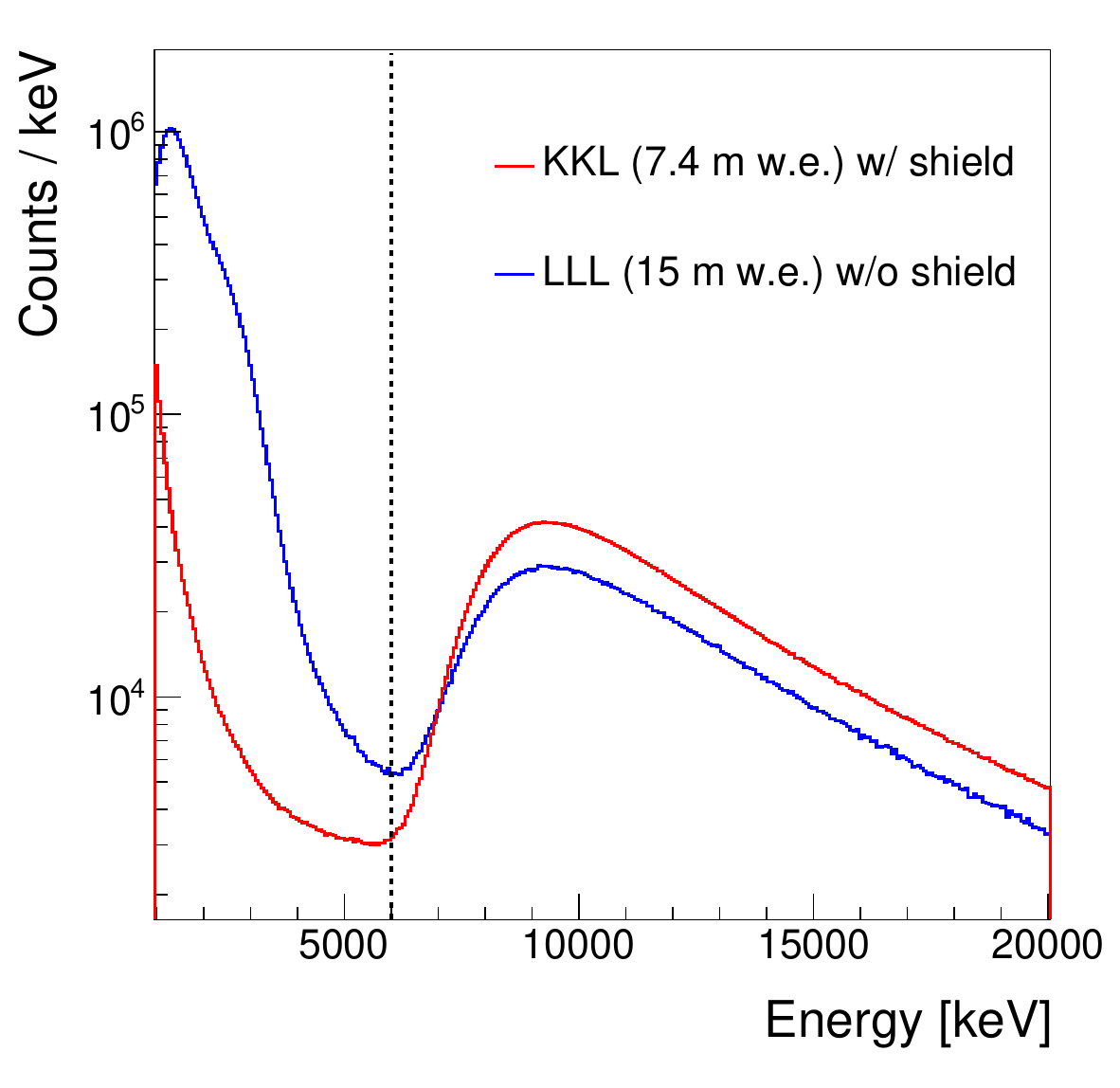}
    \caption{Scintillation light spectra of the inner muon veto top plate as measured by one PMT a) unshielded at MPIK underground lab (15 m w.e.) (blue) and b) inside the \conusplus~shield at KKL (9.2 m w.e.) (red). The 9.2 m w.e. is the estimated overburden from the combination of the average overburden in the room (7.4 m w.e.) and the shield material over the inner veto (1.8~m w.e).}
    \label{scintillator_spectra}
\end{figure}

The energy spectra measured without shield at the LLL of MPIK (blue) and inside the massive \conusplus~shield at KKL (red) with one PMT from the top plate are shown in figure~\ref{scintillator_spectra}. At lower energies the \conusplus~shield strongly reduces the $\gamma$-ray background, while for the tests at MPIK the Compton edge of the 2614.5~keV $\gamma$-rays from $^{208}$Tl is visible. At high energy a bump can be observed at $\sim$9.3~MeV, consistent with muons with a vertical incidence in the top plate. Considering the thickness of 4.2~cm of the plate, this corresponds to an average energy deposition per unit path length of 2.2~MeV cm$^{-1}$, in agreement with the values presented in~\cite{TKACZYK201896}. The high light yield of the scintillator plates provides a clear separation between $\gamma$- and $\mu$-events. The energy threshold is set individually for each PMT, maximizing the discrimination of both components (black line). The $\mu$-rate within the top layer was determined to be (40~$\pm$~1)~Hz at LLL, which corresponds to about (65~$\pm$~2)~Hz m$^{-2}$, a reduction factor of $\sim$3 with respect to surface. Considering a 20~ns PMT coincidence window, a total rate of (75~$\pm$~2)~Hz was measured for the whole inner veto, meaning that more than half of the muons are detected by the top plate. In comparison, the rate measured for the top plate and the full inner veto at KKL was (55~$\pm$~2)~Hz and (101~$\pm$~2)~Hz, respectively.    

The comparison between the energy spectra measured in both locations can give a tentative idea of the effective overburden for the suppression of the cosmic-ray muon component. The scaling factor between both locations is 1.42~$\pm$~0.03, consistent with the rate differences. The LLL has a well-known overburden of 15~m~w.e~\cite{Hakenmuller:2016fsv}, leading to an overburden of $\sim$9~m~w.e. at KKL. It is important to notice that the \conusplus~shield material over the inner veto provides an additional 1.8~m w.e. of overburden. Therefore, the determination is in agreement with the 7.4~m w.e. measurement mentioned in Sec.\,\ref{sec:2}. Further details about the muon veto performance at KKL are provided in Sec.\,\ref{sec:4.1}.
 
\subsection{Germanium detector upgrade}
\label{sec:3.2}


The five PPC HPGe detectors~\cite{Bonet:2020ntx} C1-C5 used in the former CONUS experimental activities were refurbished by the company Mirion Technologies (Canberra) in Lingolsheim, France (Mirion–Lingolsheim), in order to improve their electronic performance and specifically to obtain a once again lower energy threshold. The four detectors with the best performance, C2-C5, were installed in the \conusplus~experiment, while the fifth detector C1 remained at MPIK for auxiliary measurements.

Each HPGe diode has a height and a diameter of 62~mm~$\times$~62~mm, each corresponding to a total crystal mass of 996~g. The dimensions of the dead and transition layers were kept the same during refurbishment, providing a total active mass of 3.74~kg as target material for \conusplus. The HPGe diodes are cooled with Cryo-Pulse 5 Plus (CP5+) electrically powered pulse tube coolers in order to satisfy the reactor safety requirements. However, for \conusplus~the cryocoolers have been upgraded by replacing the 2-fan ventilation system with a water-cooled chiller system. This helps to suppress mechanical vibrations, which previously induced microphonic events in the HPGe detectors, partly in correlation with room temperature variations. Indeed, during the CONUS operation at KBR  a noise rate variation of 3-4~\% was observed for a temperature change of 1$\degree$C \cite{Bonet:2020ntx}. With the new system, the level of microphony is reduced in a wide frequency band and the noise rate temperature dependence has dropped to $\sim$0.5\%/$\degree$C. 

The p-type HPGe diodes undergo electrical depletion when a positive reverse bias voltage of several thousands volts is applied to the surrounding n+ contact. Within the diode, ionization events produce electron-hole pairs, which then drift in opposite directions to the electrodes. Holes drift towards the p+ contact, where a custom-built charge sensitive preamplifier (CSP) amplifies the signal. The custom-built junction field-effect transistor (JFET) of the original CSP design was replaced by an application-specific integrated circuit (ASIC), which has recently demonstrated a better performance in terms of noise level and detection trigger efficiency~\cite{Bonhomme:2022lcz}. In order to minimize further the electronic noise, the new design keeps a pulsed-reset instead of a resistive feedback. This requires, however, the generation of resets when the increasing baseline has reached the saturation of the dynamic range. A proper logic generates rectangular inhibit signals (TRP) to veto unwanted spurious HPGe detector signals generated shortly after the resets. Since the electronic noise depends also on the detector capacitance, that is proportional to the point-contact area, the already small p+ contact size was further reduced. Finally, for the connection between the p+ contact and the cold front-end electronics the noise-suppressing bonding technique was applied instead of the formerly used contact pins, in order to minimize the electromagnetic interference by reducing the inductive and capacitive coupling. All these modifications were successful and improved further the noise level and thus the energy resolution and energy threshold.


The detector-dependent energy resolution was evaluated for the 59.6~keV $\gamma$-line from an $^{241}$Am source calibration and for artificial signals generated by a pulse generator (Keysight 33500B) with the same pulse rise time as the physical signals. The pulses are injected through a specific circuit implemented in the HPGe pre-amplifiers. The energy resolution measured at MPIK and at the \conusplus~experimental site are reported in table~\ref{resolution} in terms of full-width at half-maximum (FWHM). The achieved pulser values are well below the design specification limit of 65~eV$_{ee}$. Furthermore, the pulser resolution is slightly improved at KKL compared to the values obtained at MPIK, despite the environmental challenges at reactor site. In general, the energy resolution has been improved by more than 15~eV$_{ee}$ (correspondingly up to 30\%) with respect to the original CONUS detectors (cmp. with Table 8 in~\cite{Bonet:2020ntx}).

\begin{table}[bht]
\setlength\extrarowheight{4pt}
\begin{center}
\begin{footnotesize}
\caption{\label{resolution} Peak resolutions of the four upgraded HPGe detectors C2-C5 used in \conusplus. The first two columns are the FWHM of artificial peaks obtained from pulser measurements at KKL and MPIK. The last column corresponds to the FWHM of the physical 59.6~keV $\gamma$-line from an $^{241}$Am source irradiation at MPIK.}
\begin{tabular}{c|cc|c}
\hline
\hline
\multicolumn{1}{l|}{Detector} &
\multicolumn{2}{c|}{Pulser~[eV$_{ee}$]} &
\multicolumn{1}{c}{$^{241}$Am~[eV$_{ee}$]}  \Tstrut\Bstrut\\ 
\hline
           &		KKL	& MPIK	& MPIK \Tstrut\Bstrut\\
\hline
C2 & 48~$\pm$~1 & 50~$\pm$~1 & 318~$\pm$~2 \\
C3  & 47~$\pm$~1 & 49~$\pm$~1 & 308~$\pm$~2 \\
C4 & 47~$\pm$~1 & 48~$\pm$~1 & 310~$\pm$~2 \\
C5 & 47~$\pm$~1 & 49~$\pm$~1 & 314~$\pm$~2 \Tstrut\Bstrut\\
\hline
\hline
 \end{tabular}
\end{footnotesize}
\end{center}
\end{table}

\begin{figure}
    \centering
    \includegraphics[width=0.48\textwidth]{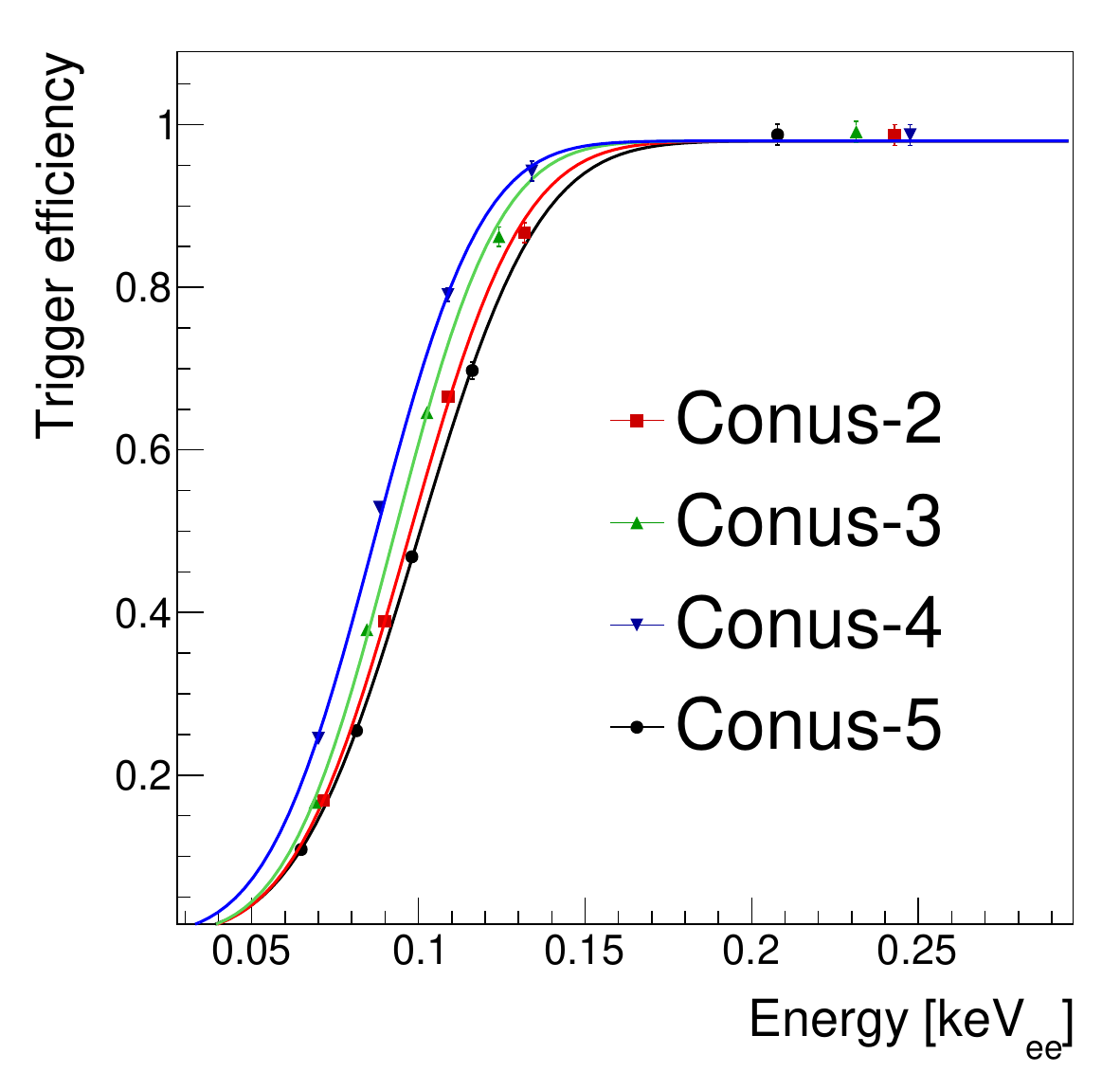}
    \caption{Trigger efficiency for physical events of the four upgraded HPGe detectors C2-C5 as measured at the \conusplus~experimental site in KKL.}
    \label{trigger_efficiency}
\end{figure}

The trigger efficiency of the four upgraded C2-C5 detectors at low energies was determined by injecting artificial signals using a pulse generator. The amplitudes of the signals were step-wise varied in order to determine the detector capability to identify physical events in the region of interest, which lies close to the noise region. The trigger efficiencies of the C2-C5 detectors measured at KKL are shown in figure~\ref{trigger_efficiency}. The experimental trigger efficiency curves can be described using the following fit function~\cite{Ackermann:2024kxo}:

\begin{equation}\label{eq:trigger_efficiency}
 \\\\ \varepsilon_{trig} = 0.5\cdot\left(1+\text{erf}\left(\frac{E_{ee}/eV_{ee}-t_1}{t_2}\right)\right)\, ,
\end{equation}
with t1 being a value in [88, 100] and t2 in [35, 40], respectively. Notably, the trigger efficiency remains over 90\% down to 140~eV$_{ee}$, indicating again a substantial improvement compared to the achieved values in the former CONUS experiment (cmp. with Fig.~3 in~\cite{Bonet:2023kob}). 

\begin{figure}
    \centering
    \includegraphics[width=0.48\textwidth]{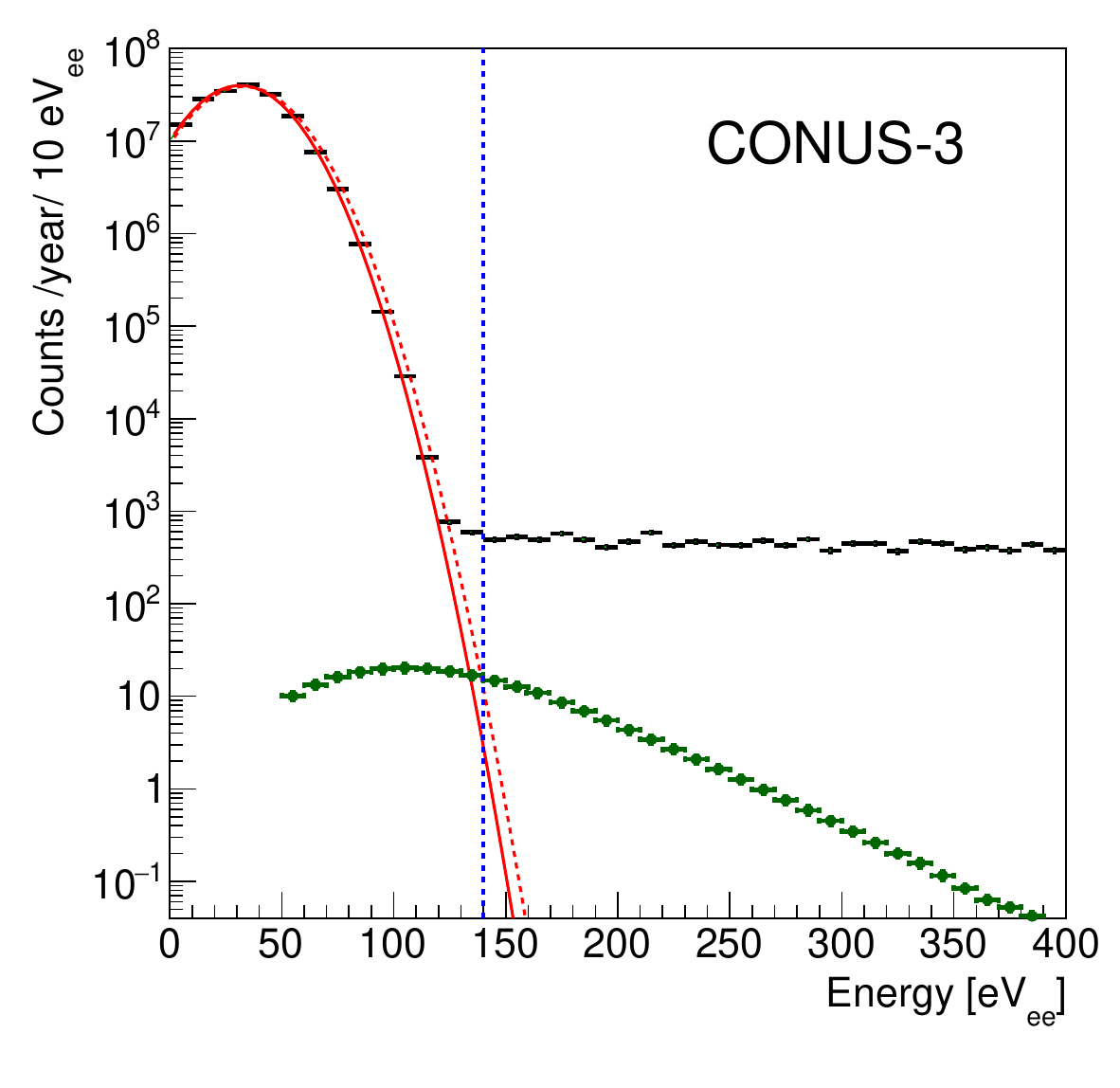}
    \caption{Background spectrum (black) measured at reactor site for the C3 detector after muon veto and TRP selection cuts in the energy region close to the noise edge. The noise can be described with a Gaussian function (red). The dashed red line accounts for the maximum accepted noise variations during the data taking. The 10\% of the expected \CEvNS~signal is indicated by green points. The estimated energy threshold of 140~eV$_{ee}$ is represented with a blue dashed vertical line.}
    \label{energy_thd}
\end{figure}

The lower energy limit for \CEvNS~data analysis is defined independently for each detector and is based on two different criteria. First, the trigger efficiency has to be at least 20\%, and second the rate of noise events in each bin must be 10 times below the signal expectation (calculated as described in Sec.\,\ref{sec:5}). The first condition is satisfied down to 70~eV$_{ee}$ for all detectors. For the second condition, the noise peak is described with a Gaussian function (red) in a typical spectrum during operation at reactor as shown in figure~\ref{energy_thd}. The FWHM of the noise peak is similar to the measured pulser resolution with values around 47~eV$_{ee}$. The dashed red line accounts for the maximum accepted noise variations during the run. The 10\% of the \CEvNS~signal prediction is represented by green points, with an energy uncertainty of 10~eV$_{ee}$. An energy threshold between 150-160~eV$_{ee}$ is determined, depending on the \conusplus~detector, improving the best threshold values achieved previously in CONUS by 50-60~eV$_{ee}$ (correspondingly by 30\%).

 
In conclusion, all these improvements of the detector performance will allow to significantly increase the number of expected \CEvNS~events in \conusplus~with respect to the former CONUS experiment (cmp. Sec.~\ref{sec:5}). A publication with more details about the HPGe detector upgrade campaign is in preparation.

\subsection{Data acquisition system (DAQ)}
\label{sec:3.3}

The \conusplus~data acquisition is performed with a combination of two CAEN digitizers with a shared synchronization clock, using pulse processing implemented at the FPGA level. The data acquisition system (DAQ) is managed via the commercially available CoMPASS software from CAEN. The signals from the HPGe detectors are connected to a 16-bit CAEN V1782 multi-channel analyzer with a sampling rate of 100~MHz. Each signal is split in two channels with different internal attenuations, providing two energy intervals at the same time, a high energy range up to 600~keV$_{ee}$ and a finer-binned low energy one up to 35~keV$_{ee}$. Inputs with a 10 $\mu$s AC coupling are used in order to get rid of the TRP-induced step-like rising baseline level. 

The trigger is generated with a combination of a slow and a fast triangular discriminator with shaping times of 0.6~$\mu$s and 1.8~$\mu$s, respectively. An independent discriminator threshold is set for each channel as low as possible to optimize the detection efficiency of physical low energy signals, while ensuring to not exceed 3\% in dead-time. This results in accepted noise trigger rates of approximately 800~Hz per detector. The energy of each event is reconstructed by a trapezoidal shaping filter with a rise time of 3~$\mu$s and a trapezoidal flat top of (0.2-0.5)~$\mu$s depending on the detector and optimized in terms of energy resolution. A trigger hold-off of 20~$\mu$s is applied to reject pile-up events.

The signals from the 40 PMTs of the $\mu$-veto system are registered at a rate of 62.5 MHz by a 12-bit CAEN V1740D digitizer. A trigger is generated with a simple leading edge algorithm, setting the threshold individually for each channel. All trigger events are saved and the $\mu$-veto rate estimation is performed in the offline analyses (sc. Sec.\,\ref{sec:4.1}). The inhibit TRP signals from the HPGe detectors are also acquired by the CAEN V1740D module. These signals are generated by the preamplifiers after the saturation of the dynamic range to eliminate spurious events following the resets of the baselines (cmp. Sec.~\ref{sec:3.2}). The overall TRP rate is in the [8, 30]~Hz range and is detector-dependent.

\subsection{Network and computing}
\label{sec:3.4}

\begin{figure*}
    \centering
    \includegraphics[width=0.95\textwidth]{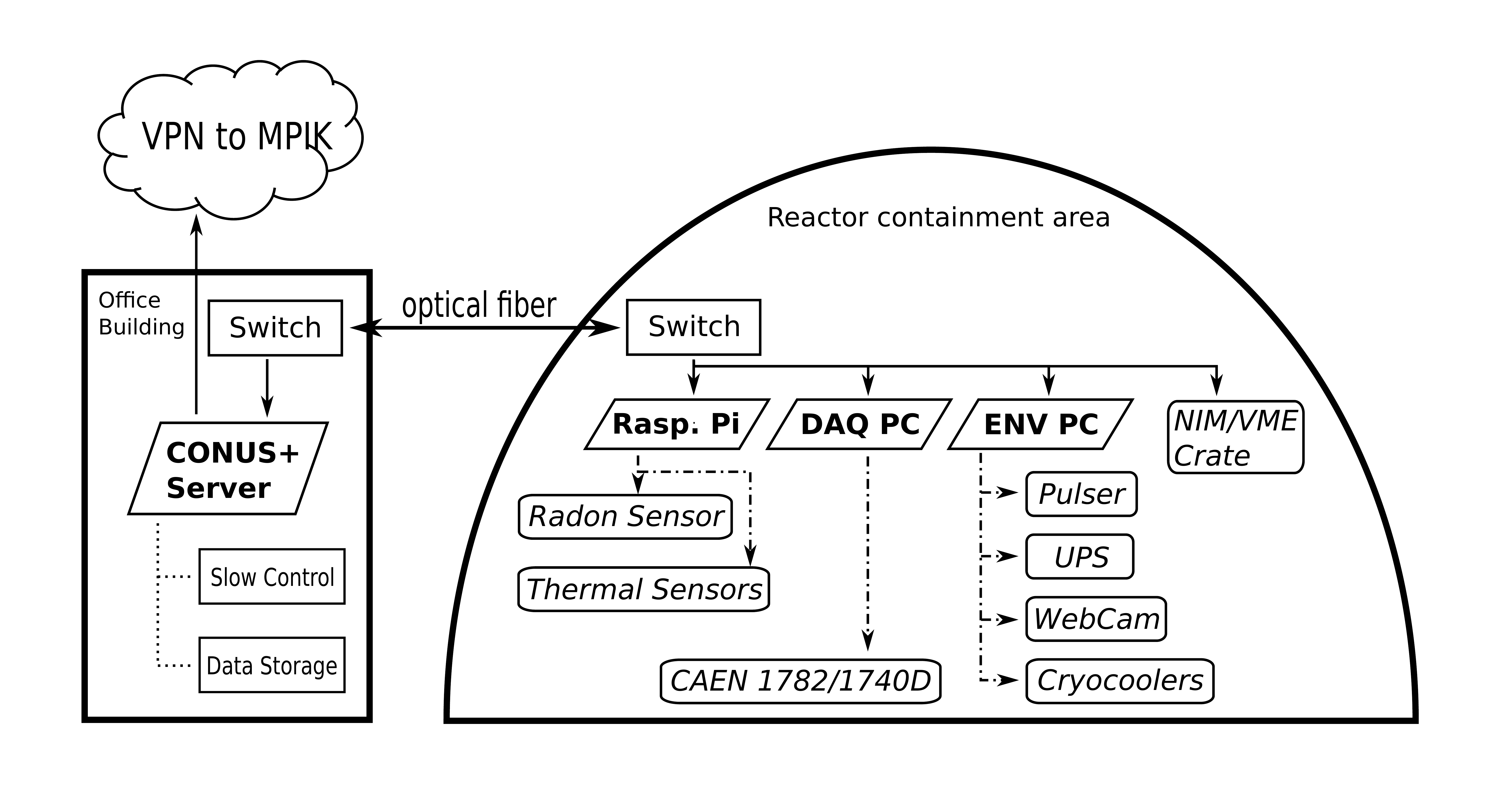}
    \caption{Scheme of the network configuration in \conusplus. A real time monitoring of the experiment is possible through a direct network connection from KKL to MPIK.}
    \label{computing_scheme}
\end{figure*}

In opposite to the situation in CONUS at KBR, \conusplus~ got the possibility to install a network at KKL. An optical fiber allows the direct connection between the \conusplus~computers in the ZA28R027 room and an external server outside the KKL containment area, which is completely independent from any IT infrastructure of the nuclear power plant. A VPN line is then established between the server and MPIK, which allows encrypted data transfer. A scheme of the full network configuration in \conusplus~is shown in figure~\ref{computing_scheme}. 

One major feature with such a network configuration is the real-time monitoring of the \conusplus~experiment, using an open-sourced InfluxDB database, forming the so-called `slow control' system. All parameters are displayed on a webUI that can be inspected remotely. Inside the containment, a Raspberry Pi drives all the environmental sensors to take temperature, humidity and Rn concentration data, while the cryocoolers, the DAQ, the high voltage modules and the UPS are directly controlled by two computers. All data are periodically uploaded to the server.  Furthermore, an alert system monitors some of the critical parameters such as the room temperature and the cryocooler power consumption. If a value exceeds a certain predefined threshold, alert messages are generated. Other non-electronic devices, as the Rn-free air flushing system, are monitored visually via a webcam. 

The network connection also makes possible the operation of the DAQ and pulser generator systems from remote and the automatic data transfer. However, the data transfer speed is limited to~5 MB/s, below the \conusplus~data production rate of 7~MB/s. For this reason, data are automatically reduced onsite, keeping only the waveform of the physics events and some noise samples coming from the HPGe detectors. A data reduction over 70\% is achieved with this procedure. The total amount of data produced every month after reduction is $\sim$1~TB. Further data processing and high-level analyses are performed on the MPIK cluster.

\section{Installation and commissioning}
\label{sec:4}

\subsection{Site preparation and detector installation}
\label{sec:4.0}

The experimental site preparation in ZA28R027 room of the KKL containment area required detailed static calculations as well as a proof of integrity for the earthquake load case. Due to the local conditions, a new support structure and a new stainless steel frame for the shield that is enforced compared to the one used at KBR, had to be designed and produced.

In April 2023, the original CONUS shield setup was dismounted at the KBR reactor and moved to the KKL nuclear power plant. Together with the upgraded \conusplus~detectors and the new shield components, i.e. the second muon veto system (cmp. Sec. \ref{sec:2}) and the new stainless steel frame, \conusplus~was set up in ZA28R027 room in July-August 2023. An additional wall was erected next to the setup to provide a separated area for \conusplus, which can be kept cleaner and at a constant temperature of about 22$\degree$C using an air cooling (AC) system. In addition, electrical and network connections independent of any other reactor system were provided (cmp. Sec.~\ref{sec:3.4}). 

\begin{figure}
    \centering
    \includegraphics[width=0.48\textwidth]{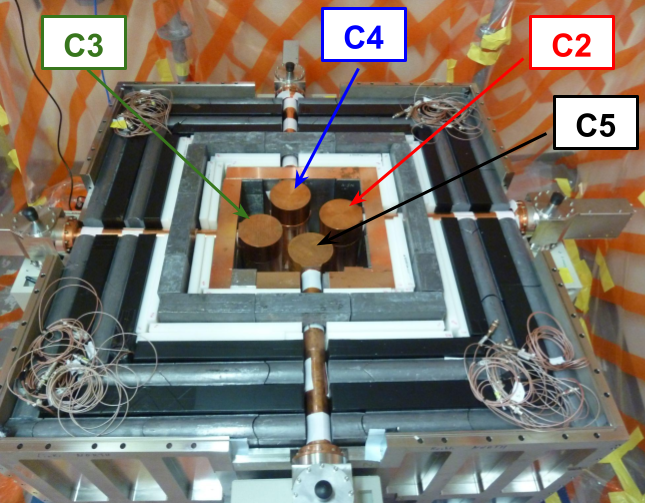}
    \caption{Photography of the four \conusplus~HPGe detectors C2-C5 within the half-opened shield during the installation in room ZA28R027 of KKL.}
    \label{conus_installation}
\end{figure}

The \conusplus~detector-shield setup with its reduced mass of 9.6 tons was assembled on the 360\,kg support structure platform (sc. figure~\ref{conus_shield}). The lower half of the shield was installed first, followed by the four HPGe detectors C2-C5 and the upper shield part. Finally, the stainless steel frame enveloping the full setup was mounted and connected to the support structure. A total installation time of $\sim$2~weeks was required. A picture of the four \conusplus~detectors inside the half installed shield inside the ZA28R027 room is shown in figure~\ref{conus_installation}. 

To be able to carry out a very low background experiment like \conusplus, it was further mandatory to preserve the radiopurity of all detector-shield components during transport and installation. Specifically, large efforts were undertaken to avoid any contamination from radioisotopes such as $^{60}$Co, $^{54}$Mn and $^{137}$Cs that can be found in dust particulates on surfaces inside the reactor building\cite{Sanchez:2024}. So, every \conusplus~element was wrapped into plastic foil after its dismantling at ZA408 room of KBR and was only unpacked directly before the assembly at ZA28R027 room of KKL. In addition, the ZA408 and ZA28R027 rooms were often cleaned and a strict cleaning protocol was followed including regular exchanges of shoe covers and gloves. Further, regular wipe tests of surfaces such as the floor and elements of the detectors and the shield were carried out at the two experimental sites. The wipe tests demonstrated that the radiopurity of the setup was preserved. 


\subsection{Commissioning of the experiment}
\label{sec:4.1}

Since \conusplus~is located at shallow depth, cosmic radiation has a large influence on the total background. Thus, an efficient suppression of this radiation type is of paramount importance for the experiment. For this reason, the performance of the $\mu$-veto system at KKL has been studied in detail. Individual energy thresholds are defined for each PMT in the muon veto system and a time coincidence window of 20~ns between PMTs. The total muon veto rate during reactor on periods is ($260\pm1$)~Hz and ($116\pm1$)~Hz for the outer and inner veto respectively, while the combination of both vetoes leads to a rate of ($274\pm1$)~Hz. Considering a 450~$\mu$s muon veto window as applied in the CONUS Run-5 analysis~\cite{Ackermann:2024kxo}, this results in a $\sim$12\% dead time. The rate for the outer and inner veto decreased during reactor off periods to ($210\pm1$)~Hz and ($112\pm1$)~Hz, while the combined rate is ($214\pm1$)~Hz. This effect was previously observed in CONUS~\cite{Bonet:2021wjw} where the muon veto rate (only outer veto layer) decreased from ($185\pm1$) to ($83\pm1$)~Hz. The smaller rate variation suggests a minor impact of reactor-correlated high-energy $\gamma$-rays in the \conusplus~muon veto system. 

\begin{figure}
    \centering
    \includegraphics[width=0.48\textwidth]{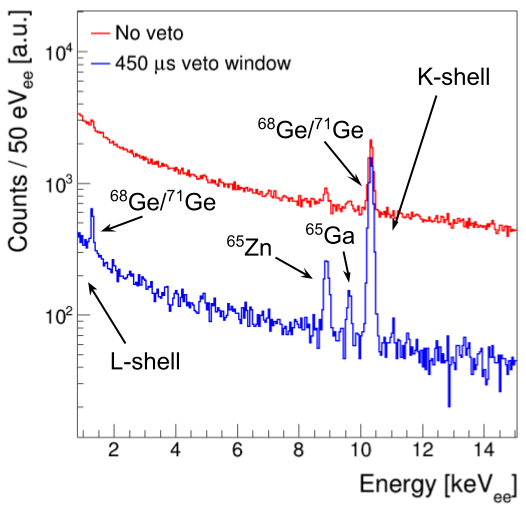}
    \caption{Energy spectra in the [0.8, 15]~keV$_{ee}$ energy range without temporal muon veto cut (red) and after applying a 450~$\mu$s wide muon veto cut (blue).}
    \label{spectra_low}
\end{figure}
\begin{figure}
    \centering
    \includegraphics[width=0.48\textwidth]{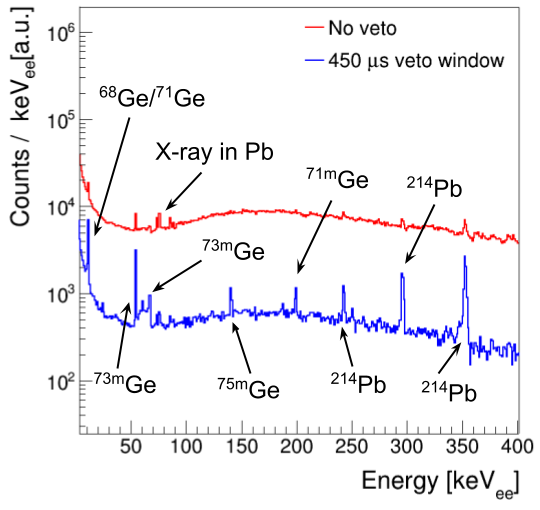}
    \caption{Energy spectra in the [5, 400]~keV$_{ee}$ energy range without temporal muon veto cut (red) and after applying a 450~$\mu$s wide muon veto cut (blue). }
    \label{spectra_high}
\end{figure}

The energy spectra in the [0.8, 15]~keV$_{ee}$  and [5, 400]~keV$_{ee}$ energy ranges are displayed in figure~\ref{spectra_low} and figure~\ref{spectra_high}, respectively, without muon veto cut (red) and after applying a 450~$\mu$s wide muon veto cut (blue). In the absence of an active muon veto system, the background is primarily dominated by prompt muon-induced events. After applying the veto, over 98\% of the background events are rejected in the [0.8, 15]~keV$_{ee}$ region, reaching the designed \conusplus~background rejection efficiency. This, together with small improvements in radiopurity due to the upgrade of the electronics, allows to achieve similar background levels like for the deeper located CONUS experiment at KBR.  

After applying the muon veto cut, several X-ray lines become visible. These correspond to the binding energies of the K shells at 9.0~keV, 9.7~keV, and 10.4~keV, as well as those of the L shells at 1.1~keV, 1.2~keV, and 1.3~keV, and originate from cosmogenic induced $^{65}$Zn, $^{68}$Ga, and $^{68}$Ge+$^{71}$Ge decays inside the HPGe diodes. Specifically, $^{71}$Ge radioisotopes are produced continuously also at shallow depth by the 1~MeV neutrons, which are released in muon interactions with the Pb of the \conusplus~shield~\cite{Hakenmuller:2016fsv}. In particular, for \conusplus~a constant rate of $\sim$35~counts~kg$^{-1}$d$^{-1}$ and $\sim$5~counts~kg$^{-1}$d$^{-1}$ are produced for the 10.4~keV and 1.3~keV X-ray lines in all detectors. Using these spectral lines and count rates, it is possible to calibrate the \conusplus~energy spectra with a precision of 10~eV$_{ee}$.


In figure~\ref{spectra_high} at higher energies the $^{71m}$Ge, $^{73m}$Ge and  $^{75m}$Ge metastable states are visible at 198.3~keV, 53.4~keV and 66.7~keV, and 139.5~keV, respectively. The 75.0~keV and 84.9~keV X-ray lines produced by muon-induced events in the Pb shield are strongly suppressed due to the operational muon veto system.

Three $\gamma$-lines are observed at 242.0~keV, 295.2~keV and 351.9~keV, which are produced by $^{214}$Pb, a daughter nuclide from the airborne $^{222}$Rn. Rn isotopes, specifically $^{222}$Rn with its longer half-life of 3.8~d, can diffuse from the environment into the HPGe detector chamber due to the fact, that Rn is an inert gas and that the steel cage enclosing the shield, even though sealed with Rn tape, lacks a full tightness. The Rn concentration in air has a significant fluctuation over time, which is caused by temperature variations within the building and alterations in the air ventilation system, particularly significant during reactor outages. These variations affect the HPGe detector background stability even at low energies~\cite{Bonet:2021wjw}, thus it is important to minimize the Rn concentration inside the detector chamber. The Rn concentration outside the shield at KKL is monitored constantly over time using a commercial device. A typical value of 100~Bq\,m$^{-3}$ and fluctuations up to 80~Bq\,m$^{-3}$ are observed. The Rn level is monitored in each detector via the integral background in the [100, 400]~keV$_{ee}$ region and with the $\gamma$-line count rate at 351.9~keV, which has the highest branching ratio compared to the other Rn-associated $\gamma$-lines. 

\begin{figure}
    \centering
    \includegraphics[width=0.48\textwidth]{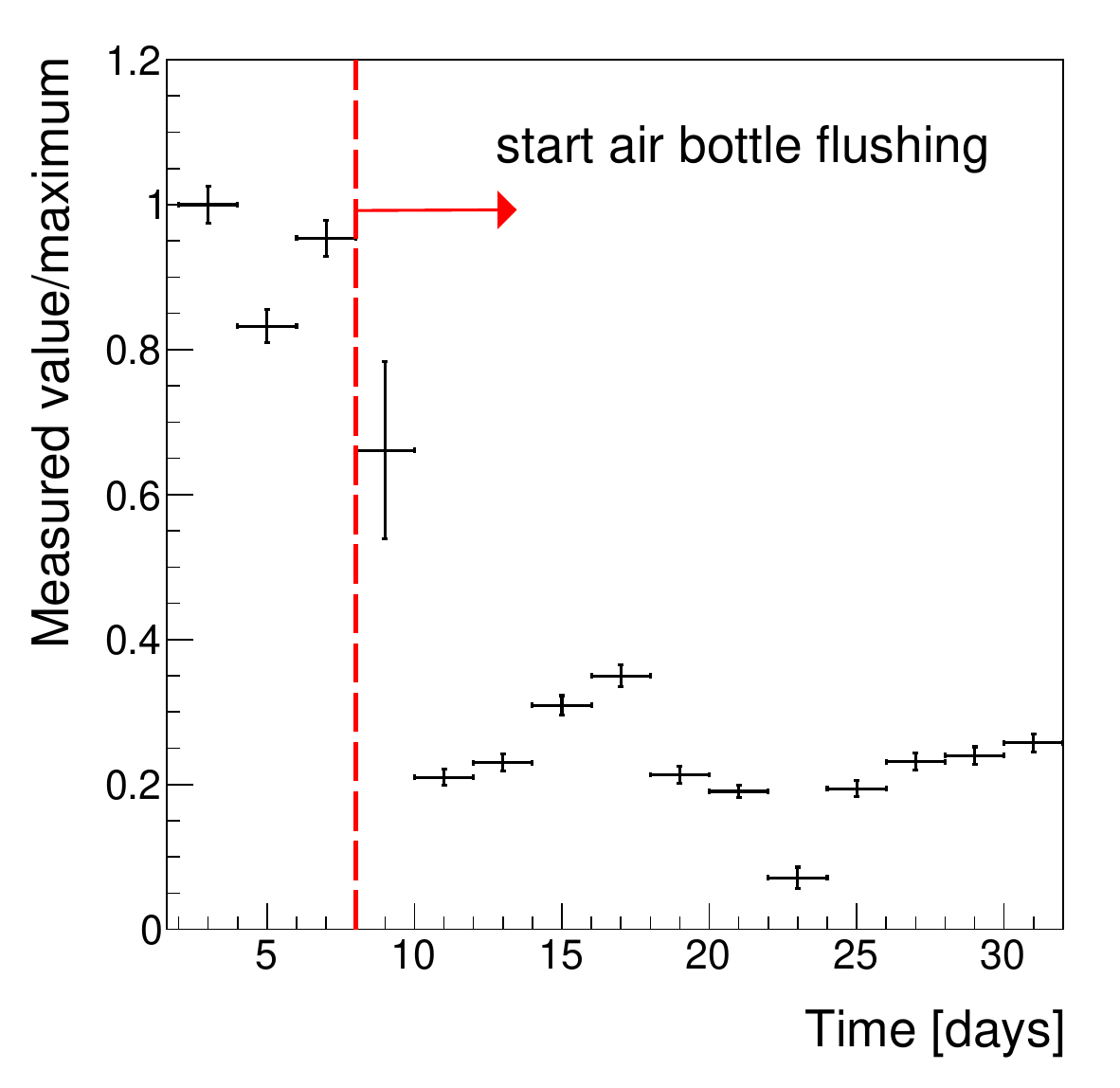}
    \caption{Evolution of the relative integral count rate of the C4 detector in
[100, 400] keV$_{ee}$ during the commissioning phase at KKL. After the start of the flushing with Rn-free air, a large rate reduction is observed.}
    \label{radon_evolution}
\end{figure}

For practical reasons, the best solution to suppress Rn at reactor site is to purge the detector chamber with breathing air from bottles, that have been stored for a minimum of three weeks, allowing sufficient time for Rn to decay away. Within \conusplus, eight 10~l bottles are connected together with a total air volume of 2.4~m$^{3}$, which allows an airflow of 1.3~standard liters per minute, while also maintaining a reasonable frequency of bottle exchange. Figure~\ref{radon_evolution} demonstrates the significantly decrease in the integral count rate upon the start of flushing, reducing the impact of Rn decays to the total background. The count rate in the 351.9~keV line decreases down to $\sim$50~counts d$^{-1}$. Using the ratio between the rate of this $\gamma$-line and the impact on the background found in \cite{Bonet:2021wjw}, a rate contribution of 2~counts~kg$^{-1}$~d$^{-1}$ in the [0.4, 1]~keV$_{ee}$ energy region is expected for the \conusplus~detectors. Future improvements of the airflow system might allow to further reduce this background.

A deep understanding of the background is crucial, especially events correlated with the reactor thermal power can be troublesome, as they can mimic \CEvNS~interactions. For this reason, two spherical Centronic SP9 proportional counters, filled with 2.3~bar of $^{3}$He and 1.2~bar of Kr were installed in the room, to continuously monitor the thermal neutron rate. One of the counters was embedded in an 8\inches~PE sphere modifying the neutron energy detection sensitivity to fast neutrons of around 1~MeV. During reactor on periods, an average rate of 0.62~neutrons s$^{-1}$ is measured with the bare counter, while a rate below 0.1~neutrons s$^{-1}$ is measured in the counter enclosed within the sphere. 

Additionally, the most significant detector and environmental parameters such as the CP5+ power consumption, the CP5+ compressor temperature, the room temperature, and humidity are continuously monitored (cmp. Sec.~\ref{sec:3.4}). The temperature in the room without the AC system has a mean value of 38$\degree$C and a variation of up to 4$\degree$C. After the installation of the AC system, an average temperature of 22$\degree$C with a variation  below 1$\degree$C was obtained. This has allowed to achieve highly stable environmental conditions and to minimize the rejected data periods due to unstable noise conditions.

\section{Signal prediction and future prospects}
\label{sec:5}

The experimental signature of a \CEvNS~interaction is a tiny nuclear recoil, whose maximum energy is inversely proportional to the mass of the struck nucleus and depends on the energy of the incoming neutrino. In the case of a Ge target and reactor antineutrinos, this corresponds to nuclear recoil energy deposits of a few keV only. Moreover, due to dissipation processes known as `quenching effect' \cite{lindhard1963range}, the ionization energy observed in the detector is even smaller than the deposited nuclear recoil energy. As the potential  \CEvNS~signal increases inversely to the deposited nuclear recoil energy, a sub-keV$_{ee}$ energy threshold is mandatory for all reactor-based \CEvNS~experiments. 

\begin{figure}
    \centering
    \includegraphics[width=0.48\textwidth]{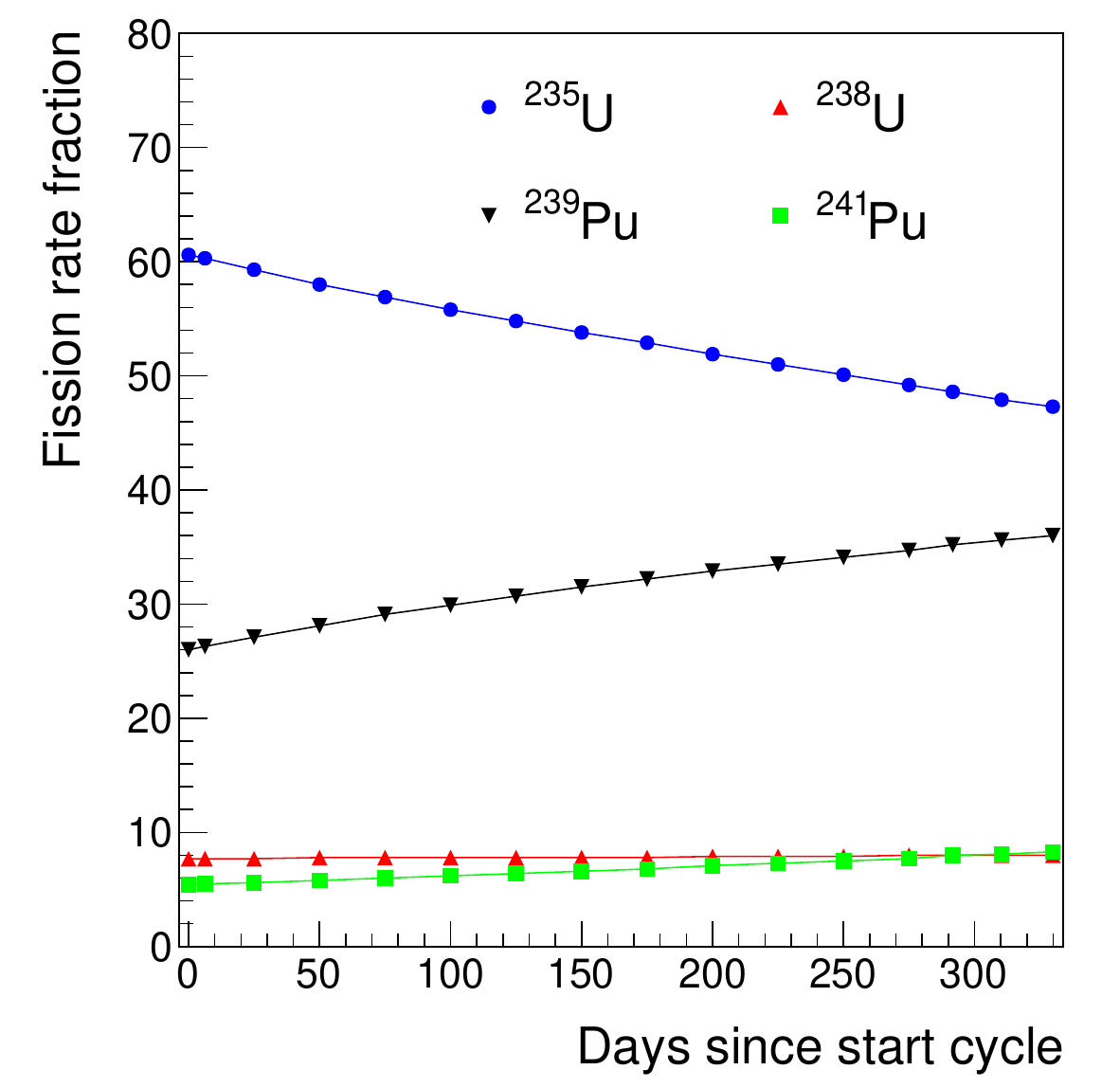}
    \caption{Exemplary average fission rate fractions of the four dominant fissile actinides as function of one typical reactor cycle evolution at KKL.}
    \label{fission_evolution}
\end{figure}

At nuclear reactors, the electron antineutrinos are produced in $\beta^{-}$ decays of neutron-rich fission fragments. The antineutrino emission depends on the combined spectra from the fission of the most important four fission nuclides present in the reactor fuel elements: $^{235}$U, $^{238}$U, $^{239}$Pu and $^{241}$Pu~\cite{Kopeikin:2012zz}. These are combined using a set of fission rate fractions to calculate the reactor antineutrino flux. The evolution of the fission fractions of the different nuclides as provided by the KKL power plant during a typical reactor cycle is shown in figure~\ref{fission_evolution}. The four mentioned actinides account together for more than 99.5\% of all fission reactions in the reactor core. The averaged values are: 53\%, 8\%, 32\% and 7\%, respectively. During one reactor cycle, the $^{235}$U is consumed decreasing from 61\% to 47\%. Pu nuclides are produced from $^{238}$U neutron captures increasing the fission fraction of $^{239}$Pu from 26\% to 36\%.

The prediction of the antineutrino spectrum is based on the method proposed in~\cite{DayaBay:2021dqj}, previously used for CONUS in~\cite{Ackermann:2024kxo}. The high energy part of the spectrum is included with the Daya Bay measurement above 8\,MeV~\cite{DayaBay:2022eyy}, while the antineutrino emission below the IBD threshold of $\sim$1.8~MeV is incorporated through the results of \cite{Estienne:2019ujo}. The reactor parameters already mentioned in Sec.~\ref{sec:2} are considered for the \conusplus~signal estimation, using the averaged values of the fission fractions and an average reactor thermal power of 3.6~GW. The measured trigger efficiency curves shown in figure~\ref{trigger_efficiency} are applied for each detector. The CONUS signal prediction is based on the values already published in~\cite{Ackermann:2024kxo}. In both cases, the quenching effect is described by the Lindhard theory~\cite{lindhard1963range} with a quenching parameter of $k=(0.162\pm0.004)$, as determined in~\cite{Bonhomme:2022lcz}. 

\begin{figure}
    \centering
    \includegraphics[width=0.48\textwidth]{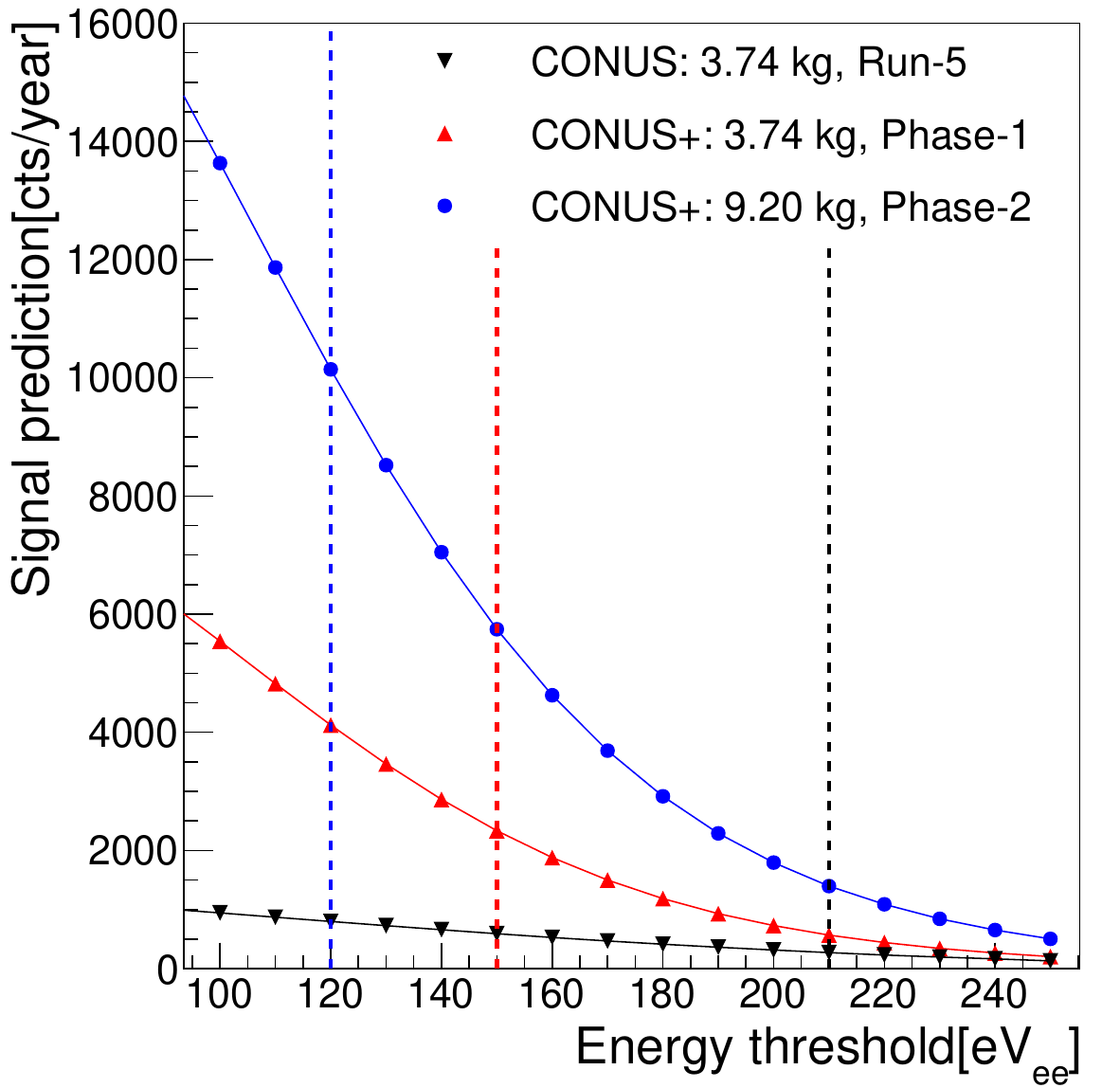}
    \caption{Expected annual integral count rate of \CEvNS~signals in the last data collection period (Run-5) of CONUS (black), Phase-1 of \conusplus~(red) and Phase-2 of \conusplus~(blue). The energy thresholds of 210~eV$_{ee}$, 150~eV$_{ee}$ and 120~eV$_{ee}$ in all three experimental phases are indicated with dashed lines. The signal prediction increases by a factor of 9 in \conusplus~Phase-1 with respect to CONUS, and it is expected to increase about 4~times more in \conusplus~Phase-2 assuming an improved energy threshold.}
    \label{signal_prediction}
\end{figure}

The number of expected \CEvNS~signal events per year is shown in figure~\ref{signal_prediction} for three cases: a) CONUS Run-5 (black), b) the ongoing Phase-1 of \conusplus~(red) and c) the future Phase-2 of \conusplus~(blue). As expected, the signal rises up exponentially with a lower energy threshold in all three cases.
The difference in the slope of the CONUS Run-5 and \conusplus~Phase-1 curves is due to the different trigger efficiency. Considering further the improved energy threshold from 210~eV$_{ee}$ to 150~eV$_{ee}$ (vertical dashed lines), the expected integral count rate in \conusplus~Phase-1 will amount to $\sim$2300~\CEvNS~per year. Although the antineutrino flux at KKL is 0.6~times smaller than at KBR, the \CEvNS~signal prediction for \conusplus~Phase-1 will increase by a factor of 9 with respect to the last CONUS Run-5. Even considering the same energy threshold of 210~eV$_{ee}$, the number of expected \CEvNS~events will still be 2.1~times larger due to the better trigger efficiency.   

Within a future Phase-2 of the \conusplus~experiment, the current four 1\,kg~HPGe detectors could be replaced by new 2.46\,kg HPGe detectors. This will provide a total crystal mass of 9.84~kg, although an active mass of $\sim$9.2~kg is considered as a conservative estimation. The detectors will fit inside the current \conusplus~shield without any modifications, providing comparable background conditions to the ongoing Phase-1 of the experiment. Moreover, preliminary studies at MPIK suggest the possibility to further decrease the energy threshold, while keeping a similar trigger efficiency as in \conusplus~Phase-1. As depicted in figure~\ref{signal_prediction}, the combination of the larger mass with an improved energy threshold could significantly boost the signal prediction for \conusplus~Phase-2, assuming an energy threshold of 120~eV$_{ee}$ an annual integral count rate over 10000 events could be reached. This will open the possibility to determine the Weinberg angle at the MeV scale and to improve the current limits on new light vector bosons as proposed in~\cite{Lindner:2024eng}.  


\section{Summary and conclusions}
\label{conclusions}

\CEvNS~is an outstanding neutrino interaction channel predicted by the Standard Model of Elementary Particles, notable for its high cross section. This will enable the novel construction of compact neutrino detectors for basic research as well as for industrial applications. Currently, there is an increasing interest worldwide to measure \CEvNS~ with reactor antineutrinos. The CONUS experiment at the German nuclear power plant in Brokdorf achieved the best reactor-based \CEvNS~detection limit so far. However, the shutdown of this power plant triggered the search for a new experimental site, which was found at the Swiss nuclear power plant in Leibstadt. 

The CONUS shield was adapted to the new background conditions, replacing one of the original lead layers by an additional plastic scintillator layer, in order to improve the muon veto efficiency. The HPGe detectors from CONUS were also upgraded, improving the trigger efficiency, energy resolution and the energy threshold. The new energy threshold is reduced by 50~eV$_{ee}$ as compared to the previous best value of CONUS, reaching a value of 160~eV$_{ee}$. Additionally, the electrically powered pulse tube coolers were optimized to reduce vibrations. A real-time monitoring of the experiment became possible through a direct network connection from KKL to MPIK.

The \conusplus~detector setup was installed at 20.7~m from the reactor core of the KKL power plant. The commissioning phase at KKL has shown a promising performance, reaching the required background rejection efficiency over 98\% at low energies and stable environmental conditions. Additionally, the airborne Rn contribution has been successfully suppressed by flushing with a Rn-free air system. After all these improvements, the expected \CEvNS~event rate in the first phase of \conusplus~is 9~times larger than in CONUS. An additional upgrade of the \conusplus~detectors with larger diodes and possible improved energy thresholds is in preparation. In this second phase the \CEvNS~signal sensitivity will be further increased by a factor 4. 

In conclusion, \conusplus~will continue the physics program of CONUS with significant improvements. The main goal of \conusplus~Phase-1 will be the measurement of \CEvNS~with reactor antineutrino for the first time, while in Phase-2 higher precision measurements will become accessible.\\

\small

\textbf{Acknowledgements}

For technical, mechanical, electronics, DAQ, IT and logistical support we thank all involved divisions and workshops
at the Max-Planck-Institut f\"ur Kernphysik in Heidelberg. In particular we thank T. Frydlewicz and J. Schreiner in transport and custom issues, T. Apfel and M. Reissfelder for technical support during the installation of the experiment. 
We thank Mirion Technologies (Canberra) in Lingolsheim for the detector upgrades and for enduring and highly professional support.  
We express our deepest gratitude to the PreussenElektra GmbH for hosting and supporting the CONUS experiment and to the Leibstadt AG for hosting and supporting the \conusplus~experiment.
For the successful preparation, installation and operation of \conusplus~we specifically thank P. Graf, H. Eltgen, P. Kaiser, R. Meili, A. Ritter and all involved departments.
We are very grateful to the engineer offices L. Baumann AG and A. Strube for the static calculations. We thank R. Bieli for providing the Leibstadt reactor thermal power data and performing the simulation of the fission fraction evolution.  
Finally, we like to thank the XENON100 Collaboration for providing the PMTs for the inner muon veto, in particular to L. Baudis, M. Galloway and S. Lindemann. The \conusplus~experiment is supported financially by the Max Planck Society (MPG).

\normalsize


\bibliographystyle{bibliostyle}
\bibliography{references}
\end{document}